\documentclass[12pt]{cernlik}
\usepackage{epsfig}
\usepackage{amssymb}

\def\be{\begin{equation}}
\def\ee{\end{equation}}
\def\bea{\begin{eqnarray}}
\def\eea{\end{eqnarray}}

\def\beq{\begin{equation}}
\def\eeq{\end{equation}}
\begin{document}
\begin{titlepage}
\title{ON UPPER LIMITS FOR GRAVITATIONAL RADIATION}

\author{P.Astone$^1$ and G. Pizzella$^2$\\
$^1$ INFN, Sezione di Roma\\
$^2$ University of Rome Tor Vergata
 and INFN, Laboratori Nazionali di Frascati\\
 P.O. Box 13, I-00044 Frascati, Italy
}
\maketitle
\baselineskip=14pt
\vskip 4cm
\begin{abstract}

A procedure with a Bayesan approach for calculating upper 
limits to gravitational wave
bursts from coincidence experiments with multiple detectors is described.

\end{abstract}
\vspace*{\stretch{2}}
\begin{flushleft}
%%	insert here the PACS number 
%%	\vskip 2cm
\end{flushleft}
\end{titlepage}
\pagestyle{plain}
\setcounter{page}2
\baselineskip=17pt
\section{Introduction}
After the initial experiments with room temperature resonant detectors, the new
generation of cryogenic gravitational wave (GW) 
antennas entered long term data
taking operation in 1990 (EXPLORER~\cite{long}), in 1991
 (ALLEGRO~\cite{alle}), in 1993 (NIOBE~\cite{NIOBE}),
in 1994 (NAUTILUS~\cite{naut}) and in 1997
(AURIGA~\cite{auri}).

Searches for coincident events between detectors have been performed.
Between EXPLORER and NAUTILUS and between EXPLORER and
NIOBE in the years 1995 and 1996 \cite{astro}.
Between ALLEGRO and EXPLORER with data recorded in 1991~\cite{ae1991}.
In both cases no significative coincidence excesses were found
and an upper limit to GW bursts was calculated ~\cite{ae1991}.

However, the upper limit determination has been done
under the hidden hypothesis that the signal-to-noise ratio (SNR)
is very large. According to theoretical estimations the signals
expected from cosmic GW sources are extremely feeble, so small that
extremely sensitive detectors are needed. In fact, according to present
knowledge, the detectors available today have not yet reached the sensitivity
to detect even a few events per year.

Thus it is important to study the problem of the upper limit determination
in the cases the SNRs of the observed $events$ are not large.
In order to do this we have to discuss our definition of $event$.

The raw data from a resonant GW detector are 
filtered with a filter matched to short bursts \cite{fast}. We describe now
in more detail the procedure used for the GW detectors of the Rome
group, EXPLORER and NAUTILUS.

After the filtering of the raw-data, $events$ are extracted as follows.
Be $x(t)$ the filtered output of the electromechanical transducer
which converts the mechanical vibrations of the bar in electrical
signals. This quantity is normalized, using the detector calibration,
such that its square gives the energy innovation $E_f$ of the oscillation
for each sample, expressed in kelvin units.
 In absence of signals, for well behaved noise due only to the thermal
motion of the bar and to the electronic noise of the amplifier,
the distribution of $x(t)$ is normal with zero mean.
The variance (average value of the square of $x(t)$)
is called $effective~temperature$ and is indicated with $T_{eff}$. The
distribution of $x(t)$ is
\be
f(x)=\frac{1}{\sqrt{2\pi T_{eff}}}e^{\frac{x^2}{2T_{eff}}}
\label{normal}
\ee
For extracting $events$ (in absence of signals the events are just
due to noise) we set a 
threshold in terms of a critical ratio defined by
\be
 CR=\frac{|x|-<|x|>}{\sigma(|x|)}=
\frac {\sqrt{SNR_f}-\sqrt {\frac {2}{\pi}}}{\sqrt {1-\frac {2}{\pi}}}
\label{creq}
\ee
where $\sigma(|x|)$ is the standard deviation of $|x|$ and we put
\be
 SNR_f=\frac{E_f}{T_{eff}}
\label{teffeq}
\ee

The threshold is set at a value CR such to obtain,
in presence of thermal and electronic noise alone,
a number of events which can be easily exchanged among
the other groups who participate to the data exchange.
For about one hundred $events$ per day the threshold corresponds to an energy
$E_t=19.5~ T_{eff}$.

We calculate now
the theoretical probability to detect a signal with a given SNR,
in presence of a well behaved Gaussian noise.
We put  $y=(s+x)^2$ where
$s\equiv \sqrt{SNR}$ is the signal we look for and $x$ is the gaussian
noise. We obtain easily \cite{papoulis}
\be
probability(SNR)=\int_{SNR_t}^{\infty} \frac {1}{\sqrt{2 \pi y}}
e^{-\frac{(SNR+y)}{2}}
cosh(\sqrt{y\cdot SNR})dy
\label{papou}
\ee
We put $SNR_t=19.5$ for the present EXPLORER and NAUTILUS detectors.

\section{Upper limit determination}

We consider M detectors and search for M-fold coincidences over a
total period of time $t_m$ during which all detectors are
in operation. Be $\bar{n}$ the average number of accidental
coincidences (due to chance) and $n_c$ the number of coincidences
which are found within a given time window.

For events which have a Poissonian distribution in time the expected
average number of M-fold accidental coincidences is given \cite{libro} by
\be
\bar{n}=Mw^{M-1}\prod^{1,M}_kn_k
\label{enne}
\ee
where $n_k$ is the event density of the $k^{th}$ detector.

The accidental coincidence distribution can be estimated experimentally
by proper shifting \cite{weber} the event occurrence times of each
detector. In the case of Poissonian distribution the 
average number of the M-fold accidental coincidences coincides
with that given by eq. \ref{enne}.
The comparison between $n_c$ and $\bar{n}$ allows to reach some
conclusion about the detection of GW
or to establish an upper limit to their existence.

In paper \cite{ae1991} and in the previous paper \cite{amaldi}
the upper limit has been estimated as follows. It has been found that,
for various energy levels of the observed events, the number 
$n_c$ was smaller than or did not exceeded significantly $\bar{n}$.
Such numbers $n_c$, one for each energy level, were used for calculating
the upper limit. A Poissonian distribution of the number of the
observed events was considered together with the hypothesis of an 
isotropic distribution in the sky of the GW sources.
The value of $h$ (adimensional perturbation of the metric tensor)
was then derived $from$ the energy levels, using
the detector cross-section for gravitational waves.

This procedure can be objected on two points:\\
a)The most important point is that, as shown in \cite{snr}, for SNR small
and up to values of a few dozens, the energy of
an event is $not$ the energy of the GW absorbed
by the detector. This means that we cannot deduce the value of $h$
directly from the energy levels of the observed events;\\
b)In addition, the efficiency of detection, again for SNR values up to one or
two dozens,
is rather smaller than unity, and this changes the upper limit,
particularly at small SNR.

We introduce a new procedure for estimating the upper limit,
which circumvents the difficulties indicated in the above two points.

The problem to determine the upper limit has been discussed in
several papers. In particular in paper
\cite{uppergenerale}, as indicated by the PDG,
\cite{PDG} and, more recently, in paper \cite{pia}.
According to \cite{pia} the upper limit can be calculated using the
relative belief updating ratio
\be
R(n_{GW},n_c,\bar{n})=e^{-n_{GW}}(1+\frac{n_{GW}}{\bar{n}})^{n_c}
\label{belief}
\ee
referring to a given period $t_m$ of data taking.
This function is proportional to the likelihood and it allows to
infer the probability to have $n_{GW}$ signals for given priors
(using the Bayes's theorem). It has already been used in High Energy Physics
\cite{zeus,higgs}.

We calculate the upper limit by solving the equation
\be
R(n_{GW},n_c,\bar{n})=0.05
\label{cinque}
\ee
We remark that 5\% does not represent a probability but it is an useful way
to put a limit independently on the priors\footnote[1]{
To avoid confusion we shall continue to use the words $upper~limit$,
although it would be more appropriate to call it
$standard~sensitivity~bound$ \cite{cern}.}.

Eq. \ref{cinque} has a very interesting solution. Putting $n_c=0$ we
find $n_{GW}=2.99$, independent on the value of the background
$\bar{n}$.
If we use the calculations of ref. \cite{uppergenerale} we find that,
for $n_c=0$ and $\bar{n}=0$, the upper limit is 3.09 (almost identical
to the previous one) $but$ it
decreases for increasing $\bar{n}$. The reason for this different behavior
 is due to the non-
Bayesan character of  the calculations made in \cite{uppergenerale},
as we discuss in the following.

Suppose we have $n_c=0$ and $\bar{n}\not=0$. This certainly means that
the number of accidentals, whose average value can be determined
with any desired accuracy, has undergone a fluctuation.
For larger $\bar{n}$ values, smaller is the ($a~priori$) probability
that such fluctuation occur.
 Thus one could reason that it is less likely that a
number $n_{GW}$ be associated to a large value of  $\bar{n}$, since the
 observation gave $n_c=0$. 

According to
the Bayesan approach instead, as discussed in \cite{pia}, one cannot
ignore the fact that the observation $n_c=0$ had already being made at the time
the estimation of the upper limit is considered. The Bayesan  approach requires
that, given $n_c=0$ and $\bar{n}\not=0$, one 
evaluate the $chance$ that  a number
$n_{GW}$ of signals exist. This $chance$ of a possible signal
is referred to  the observation already
made and, rather obviously, it cannot depend on the previous fluctuation
of the background, since the presence of a signal cannot be
related to the background due to the detector.
 Mathematically, it is easy to demonstrate,
using the results obtained in \cite{pia},
that due to the Poissonian
character of the number of accidentals this $relative ~chance$ 
(for $n_c=0$) is indeed independent on $\bar{n}$.

It can be seen, comparing the results of \cite{uppergenerale} with
those of \cite{pia}, that the Bayesan upper limits are
for all values of $n_c$ and $\bar{n}$
(except $n_c=\bar{n}=0$), greater than those obtained
with the non-Bayesan procedure.
In our opinion the Bayesan approach has to be preferred, and so we do in this
paper.

\begin{figure}[t]
\begin{center}
\epsfig{file=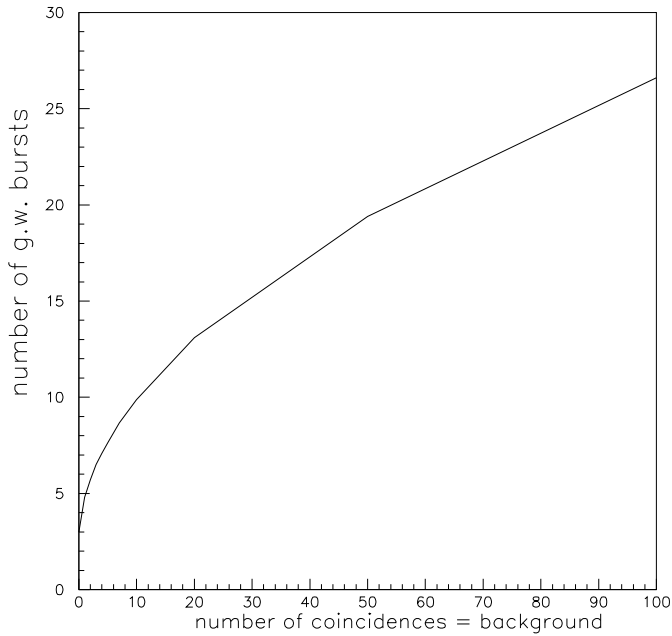,clip=,width=0.6\linewidth} 
\end{center}
%% \vspace{9.0cm}
%%\special{psfile=bayes.eps  hscale=60 vscale=60 angle=0}
 \caption{
Number of GW signals expected for the sensitivity limit of 5\%
versus the number of coincidences equal to the average number of
accidentals.
        \label{bayes} }
\end{figure}

If we have $n_c\not=0$ then we apply eq. \ref{belief}. It is interesting to show
the result for the case $n_c=\bar{n}\not=0$ for the standard sensitivity
bound of 5\%. The result is given in fig.\ref{bayes}
We note that for $n_c=\bar{n}$ and $n_{GW}<<\bar{n}$
eq. \ref{belief} can be approximated with
\be
n_{GW}\approx\sqrt{6~\bar{n}}
\label{appros}
\ee

From the result shown  in fig.\ref{bayes} it appears evident
that the lowest upper
limit is 
obtained for $n_c\sim\bar{n}\sim0$. In order to obtain $\bar{n}\sim0$ one can
raise the threshold used for determining the events. However in doing this one
diminish the efficiency of detection, as shown in eq.\ref{papou}. Whether the
procedure to raise the threshold is convenient or not, it depends
on the numerical
effects of the two competing operations. Certainly for large GW signals, when
the detection efficiency is always unity, it is much better to have a threshold
that gives $\bar{n}=0$. For smaller signals one has to consider specific cases.
However it can be seen that in the most interesting cases it is better to
raise the threshold until we get $\bar{n}\sim 0$.
 This will be shown in the section
where we reconsider the upper limit obtained with ALLEGRO and EXPLORER in
1991 \cite{ae1991}.

In the estimation of the upper limit we consider the
efficiency of detection, which we indicate with
$ \epsilon_k(SNR)$
where $k$ refers to the $k^{th}$ detector.
For EXPLORER  and NAUTILUS the theoretical efficiency is obtained from
eq. \ref{papou}.

We must relate the $h$ values of the GW to the energy $E$
absorbed by the detectors.
We have to consider that the absorbed energy depends on the direction
of the impinging GW and on its polarization. For taking care of
the various polarization we use the average value dividing the
cross section by a factor of two. We then have
\cite{australia}
\be
h=1.13~10^{-17}\sqrt{E}
\label{pola}
\ee
with the energy $E$ expressed in kelvin unit.
This formula is valid only if the GW arrives perpendicularly to the 
detector axis ($\theta=90^o$).
For a given direction we calculate the absorbed
energy using the $sin(\theta)^4$ dependency. 
We also consider that for an isotropic distribution of sources the number of
possible GW impinging directions is proportional to
$sin(\theta)^2$.

The procedure for calculating the upper limit 
is accomplished thru the following points:\\
a) consider various values of $h$;\\
b) assume an isotropic distribution of the GW sources;\\
c) for each direction $\theta$ and for each $h$ calculate the absorbed energy
$E(\theta)$ by means of eq. \ref{pola} and the $sin^4(\theta)$
dependency;\\
d) for each detector calculate the SNR for the adsorbed energy by taking
into consideration the noise $T_{eff,k}$: 
\be
SNR_k(\theta)=\frac{E(\theta)}{T_{eff,k}},~~~~~ k=1,..,M
\label{snrt}
\ee
e) using the individual efficiencies $\epsilon_k(SNR_k(\theta))$ consider the
total efficiency $\epsilon_t(\theta)=\prod_k^{1,M}\epsilon_k(SNR_k(\theta))$;\\
f) integrate $\epsilon_t(\theta)$ over $\theta$ 
with the weight $sin^2(\theta)$, because
of the assumed isotropic distribution of the sources;\\
g) from eq.\ref{belief}, given $n_c$ and $\bar{n}$, we obtain $n_{GW}$.
We then divide $n_{GW}$ by the result of point f) and obtain 
for each value of $h$ the upper limit during the measuring time $t_m$.

We remark that in this case we have not used the energy of the
observed events, as done instead previously \cite{amaldi,ae1991}.

The total efficiency is calculated with the following eq. \ref{etotale}.
\be
\epsilon_{tot}(h)=\frac{\int_0^{\frac{\pi}{2}}\prod_k^{1,M}
\epsilon_k(SNR_k(\theta))
sin^2(\theta)d\theta}{\frac{\pi}{4}}
\label{etotale}
\ee

For more clarity we show in table \ref{procedura} some of the steps
needed for our calculation, using two parallel detectors and $n_c=0$.
We use the efficiency given by eq. \ref{papou}, valid
for a well behaved noise\footnote[2]
{The real data often show a non gaussian behaviour. In this case
the efficiency differs from the theoretical one given by
eq.\ref{papou}, but one can easily make use of the efficiency experimentally
measured.}.
\begin{table}
\centering
\caption{
Procedure for calculating the upper limit with two detectors. We assume
that one
detector has noise $T_{eff}=1~mK$, the other one has noise
$T_{eff}=2~mK$. For each value of $h$
we give: maximum energy adsorbed by the detector (for $sin^4(\theta)=1$),
SNR and efficiency of detection for each detector, total weighted efficiency
(having considered an isotropic distribution of the GW sources.
Due to the angular
 weighting $\epsilon_{total}<\epsilon_{A}\epsilon_{B}$).
The upper limit is given by $\frac{2.99}{\epsilon_{total}}$.
}

\vskip 0.1 in
\begin{tabular}{|c||c|cc|cc|c|c|}
\hline
h&$E_{abs}$&Detector A&&Detector B&&&upper limit\\
$10^{18}$&$[mK]$&$SNR_A$&$\epsilon_A$&$SNR_B$&$\epsilon_B$&
$\epsilon_{tot}$&\\
&&&\%&&\%&\%&\\
\hline
2&31&31&0.88&15.5&0.32&0.12&16\\
3&70&70&1&35&0.93&0.56&4.3\\
4&126&126&1&63&1&0.76&3.6\\
10&783&783&1&392&1&0.95&3.1\\
\hline
\end{tabular}
\label{procedura}
\end{table}

\section{Ricalculation of the upper limit with the data of
ALLEGRO and EXPLORER in 1991}

In a previous paper \cite{ae1991} the upper limit for GW bursts was calculated,
using the data recorded by ALLEGRO and EXPLORER in 1991. We wish now to
recalculate the upper limit according the considerations discussed
in this paper.

\begin{figure}[t,b]
\vspace{9.0cm}
\begin{center}
\includegraphics{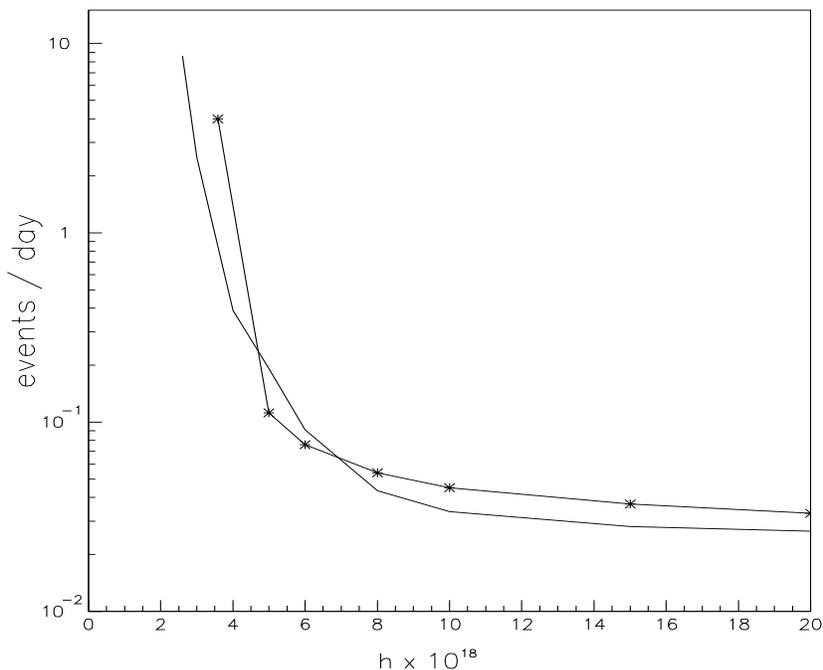}
\end{center}
\caption{
The asterisks indicate the upper limit calculated in \cite{ae1991}.
The other line indicates the upper limit evaluated with the
Bayesan approach.
	\label{super} }
\end{figure}

In 1991 the EXPLORER data filtering was done differently from that
described in this Introduction.
For both ALLEGRO and EXPLORER
the output of the electromechanical transducer was sent to lock-ins
referred to the frequencies of the resonant modes. Then
the outputs of the lock-ins (in phase and in quadrature) were filtered
searching for delta-like signals and
combined for obtaining the energy innovation, which we still indicate
with $E_f$. In this case the probability to have an event 
(above threshold $SNR_t$) due to
a signal with given SNR is obtained (see ref. \cite{alle,snr}) with 
the following equation:
\be
probability(SNR)=\int_{SNR_t}^{\infty}
e^{-(SNR+y)}
I_o(2\sqrt{y\cdot SNR})dy
\label{bessel}
\ee
Here $y=\frac{E_f}{T_{eff}}$, 
$I_o$ is the modified Bessel function of order zero, and the
noise temperature $T_{eff}$ is the average value of the
energy innovation $E_f$.

 We recall that in a time period of 123 days 70 coincidences
were found with a background of 59.3. For extracting the events the ALLEGRO
threshold was $SNR_t=11.5$ with a noise temperature
$T_{eff}\sim8~mK$. For EXPLORER the threshold was $SNR_t=10$ also with
$T_{eff}\sim8~mK$. Applying eq.\ref{belief} we find an upper limit of 
$n_{GW}=37$ over the 123 days.

According to the previous considerations we can raise the event threshold,
say for EXPLORER, in order to reduce the number of accidentals.
For instance, for
a threshold $SNR_t=24$ we get $n_c=1$ and $\bar{n}=0.74$, obtaining,
from eq. \ref{belief}, the value $n_{GW}=4.8$.

Thus the procedure for calculating the upper limit with the Bayesan
approach
when we have data at various thresholds, including cases
with $n_c$ and $\bar{n}$ different from zero, is the following.

Start with $n_c$ and $\bar{n}$ for various thresholds and use eq.\ref{belief}
for obtaining $n_{GW}$ at each threshold.
Calculate the upper limit for various values of $h$ as shown in the previous
section. For each $h$ take as upper limit
the smallest value among those obtained by varying the threshold.
Clearly at large $h$ values, when we get $n_c=0$, the upper limit is,
for the entire period of time, $n_{GW}=2.99$.

The result is shown in fig.\ref{super}
together with that obtained previously in \cite{ae1991}.
It turns out that the two upper limits are similar.

 The reason for this
is due to the fact that in applying the previous algorithm \cite{ae1991}
we started from an energy level higher than the largest energy 
of the detected (accidental) coincidences, thus obtaining, at this level
($n_c=\bar{n}=0$)
an upper limit of 3.09 very close to the value 2.99 obtained
with the Bayesan approach.
 The similarity of the results at lower $h$ values
is accidental. In the previous algorithm the increase at lower
$h$ is due only to the increase of the number $\bar{n}$ of accidentals.
In the
present algorithm the increase is due to the smaller efficiency of
detection and to the increase in $n_{GW}$ which roughly goes with
$\sqrt{\bar{n}}$ (eq.\ref{appros}).

In spite of the similar numerical results,
we believe that the procedure proposed here 
which does not extract the value of $h$ from the
energy levels of the accidental coincidences and it
uses the Bayesan approach is methodologically more correct.

\section{Discussion}
The best upper limit which can be obtained with an array of
$M$ identical parallel detectors in $M^{pl}$ coincidence cannot
go below the value 2.99,
because this is the upper limit
\cite{pia} when one finds $zero$ coincidences independently
on the background.

The basic advantage in using many detectors comes from the fact that
with many detectors it is easier to obtain
$\bar{n}\sim0$, and thus (in absence of GW) $n_c=0$. Because of the Poisson
distributions,
 the average number of accidental coincidences for M detectors in a
time window $\pm w$ is given by eq.\ref{enne}.
On the time scale of 1 second (w=1 s) it turns out
that  $n_k<<1$.
By increasing the number of detectors one obtains smaller values of
 $\bar{n}$, thus
approaching the requirement to have $n_c=0$ and then the lowest possible
upper limit.

This is certainly true at large $h$ values, where the detection
efficiency for all detectors is unity. The result, as shown in
fig.\ref{super}, is a plateau.
Instead it might be convenient at low $h$ values to use the two
most sensitive detectors, in order to have the largest possible
efficiency of detection.
The overall upper limit is then obtained by taking the smallest
ones among the values of the various upper limit determinations.

The above procedure
can be easily adjusted to the more general case of any distribution
of the GW sources, and of non-parallel detectors. 

\section{Acknowledgements}
We have benefited from useful discussions with P.Bonifazi,
G. D' Agostini and F. Ronga.

%
%\section{References}
%

%
\end{document}